\documentclass{aa}

\usepackage{graphicx}
\usepackage{txfonts}
\usepackage{wasysym}

\def\pp#1#2{\frac{\partial#1}{\partial#2}}

\begin{document}

   \title{Hydrodynamic simulations of captured protoatmospheres around Earth-like planets}

   \author{Alexander St{\"o}kl \inst{1}
      \and Ernst Dorfi \inst{1}
      \and Helmut Lammer\inst{2}}

   \institute{Institute for Astronomy, University of Vienna, T\"urkenschanzstrasse 17, A-1180 Vienna, Austria
         \and Space Research Institute, Austrian Academy of Sciences, Schmiedlstr. 6, A-8042, Graz, Austria}

\authorrunning{A. St{\"o}kl et al.}
\titlerunning{Hydrodynamic simulations of protoatmospheres}

\offprints{A. St{\"o}kl}

\date{Received ...; accepted ...}

  \abstract
   {Young terrestrial planets, when they are still embedded in a circumstellar disk, accumulate an atmosphere of nebula gas.
   The evolution and eventual evaporation of the protoplanetary disk affect the structure and dynamics of the planetary atmosphere.
   These processes, combined with other mass loss mechanisms, such as thermal escape driven by extreme ultraviolet and soft X-ray radiation (XUV)
   from the young host star, determine how much of the primary atmosphere, if anything at all, survives into later stages of planetary evolution.}
   {Our aim is to explore the structure and
   the dynamic outflow processes of nebula-accreted atmospheres
   in dependency on changes in the planetary environment.}
   {We integrate stationary hydrostatic models and perform time-dependent dynamical simulations to investigate
   the effect of a changing nebula environment on the atmospheric structure
   and the timescales on which the protoatmosphere reacts to these changes.}
   {We find that the behavior of the atmospheres strongly depends on the mass of the planetary core.
    For planets of about Mars-mass the atmospheric structure, and in particular the atmospheric mass, changes drastically and on very short timescales
    whereas atmospheres around higher mass planets are much more robust and inert.}
   {}


   \maketitle

%%%%%%%%%%%%%%%%%%%%%%%%%%%%%%%%%%%%%%%%%%%%%%%%%%%%%%%%%%%%%%%%%%%%%%%%%%%%%%%%%%%%%%%%%%%%%%%%%%%%
\section{Introduction}
%%%%%%%%%%%%%%%%%%%%%%%%%%%%%%%%%%%%%%%%%%%%%%%%%%%%%%%%%%%%%%%%%%%%%%%%%%%%%%%%%%%%%%%%%%%%%%%%%%%%

At early stages in the evolution of planetary systems, newly formed protoplanetary cores composed of solid material
orbit the host star still embedded in the circumstellar disk. These cores interact gravitationally with the
disk gas and thus accumulate an atmosphere, which at its outer margins blends into the surrounding disk
nebula environment.
For more massive cores, more substantial atmospheres form accordingly,
and, if one follows the core-accretion paradigm of gas-planet formation \citep{Perri1974, Mizuno1980},
this finally leads, for sufficiently massive cores,
to runaway accretion and the subsequent formation of a gas giant.

For the numerical study of embedded planetary atmospheres,
it is necessary to adopt some assumptions and simplifications.
These include:

\paragraph{1) Spherical symmetry:}
First, the Hill radius defined as
\begin{equation}
\label{EQ_HillRad}
R_{\rm Hill} = \left( \frac{M_{\rm pl} }{3 M_{\star}}\right)^{1/3} a \;,
\end{equation}
with $a$ the semimajor axis of the planet, gives an approximation for the sphere where
the gravitational potential
is dominated by the planet with a mass $M_{\rm pl}$ over the host star with a mass $M_\star$.
Another limit comes from the requirement that
the disk medium is essentially uniform along the perimeter of the planetary atmosphere.
The vertical structure of the protoplanetary disk can be approximated by a hydrostatic stratification
with a pressure scale height $H_p$.
Assuming a vertically isothermal disk with a temperature $T$, the pressure scale height can be
expressed as
\begin{equation}
\label{EQ_H_p}
H_p = \sqrt{ \frac{\mathcal{R} T \, a^3}{G M_\star} }
\end{equation}
where $\mathcal{R}$ is the specific gas constant, and $G$ is the gravitational constant.
Combining these two arguments, spherical symmetry is a good approximation as long as $r \ll R_{\rm Hill}$
and $r \ll H_p$. In this paper, we always assume that $R_{\rm Hill} \apprle H_p$.
Spherical symmetry also implies that rotational momentum of the atmosphere with respect to the planetary core can be neglected.

\paragraph{2) Stationary equilibrium:}
Dynamical, i.e.\ hydrostatic, equilibrium, is certainly well fulfilled during most phases of atmospheric evolution.
Thermal equilibrium, however, is much harder to justify and only adopted pragmatically
to make the problem numerically tractable.
The assumption of thermal equilibrium implies that the luminosity in the atmosphere is radially constant,
therefore the luminosity at the surface of the core is established as a main constituting parameter.
Traditionally, \citep{Hayashi1979} this luminosity has been associated with the energy released
by accreted planetesimals (assumed as continuous stream) as they impact on the surface of the core.
Equating the released energy with the gravitational energy of the stream of planetesimals,
the planetary luminosity $L_{\rm pl} $ is directly related to the accretion rate $\dot{M}_{\rm acc}$ of the planetary core
\begin{equation}
\label{EQ_L_acc}
L_{\rm pl} \simeq G M_{\rm pl} \dot{M}_{\rm acc} \left( \frac{1}{R_{\rm pl}} - \frac{1}{R_{\rm Hill}}\right) \;,
\end{equation}
where $M_{\rm pl}$ and $R_{\rm pl}$ are the mass and the radius of the planetary core, respectively.
The planetesimals are assumed to travel through the atmosphere without energy losses so that all
of the gained gravitational energy is released on the planetary surface.

Considering these two assumptions, the structure of a planetary atmosphere is described by the
usual stellar structure equations
that can be solved readily by established numerical methods.
Such hydrostatic and spherical symmetric models of primary hydrogen atmospheres around terrestrial planets have been calculated
by \citet{Hayashi1979} and \citet{Nakazawa1985}
using a constant and an analytically prescribed dust opacity, respectively.
More recently, \citet{Ikoma2006} improved upon these earlier results with the adoption of
realistic equation of state and opacities.
Particularly important for the atmospheric structure is detailed dust opacity data,
as the evaporation/condensation of various dust species
directly manifests in a sequence of atmospheric dust layers.

%Content of paper
This work deals with two distinct numerical methods: stationary modeling and time-dependent simulations.
After introducing the physical scenario in Sect.~\ref{SECT_scenario},
the method and results of stationary modeling are presented in Sect.~\ref{SECT_stationary}.
Then Sect.~\ref{SECT_dynamic} describes the time-dependent scheme and the results of the dynamic simulations.
Finally, in Sect.~\ref{SECT_Discussion}, we discuss both the effect of numerical parameters and
the astrophysical relevance of the results as well as the implications for the evolution of habitable planets.

%%%%%%%%%%%%%%%%%%%%%%%%%%%%%%%%%%%%%%%%%%%%%%%%%%%%%%%%%%%%%%%%%%%%%%%%%%%%%%%%%%%%%%%%%%%%%%%%%%%%
\subsection{Physical scenario and boundary conditions}
\label{SECT_scenario}

From statistics of IR-excess observations \citep{Haisch2001, Hillenbrand2005}
it is readily evident that the circumstellar disks
have a typical lifetime of a few Myr.
After about 10 Myr almost no protoplanetary disks are left.
The mechanism of disk evaporation is still not well understood,
though some plausible scenarios have been devised \citep[e.g.,][]{Hollenbach1994, Johnstone1998, Clarke2001, Scally2001}.
It is, however, clear that the disappearance of the nebula environment will have
significant consequences for the formerly embedded planetary atmospheres.

In this study, we focus on the direct hydrodynamical effect of the disk evaporation on the atmosphere,
ignoring many potentially important physical processes resulting from the exposure of the planetary atmosphere
to the radiation field of the host star, such as escape from a thermosphere heated
by the high extreme ultraviolet and soft X-ray luminosity (XUV) of young stars \citep{Yelle2004, Koskinen2013, Lammer2013book, Erkaev2013, Lammer2013, Lammer2014}.
The interaction of an evaporating disk with the planetary atmosphere around a cooling planetary
core recently has been studied by \citet{Ikoma2012}.
Whereas this study considers the dynamic reaction
of atmospheres, which initially are both in dynamic and thermal equilibrium, \citet{Ikoma2012} use quasi-static calculations and
cover the atmospheric evolution as the atmosphere and the solid core cool down from an initially hot state.

In terms of hydrostatic balance, the
disk nebula environment establishes a fixed gas pressure at the top of the planetary atmosphere.
Once the circumstellar disk vanishes, this gas pressure cap is released to a large degree.
In our models in particular, we reduce the gas density at the outer boundary by 15 orders of magnitude.
Starting from an initial density of $\rho = 5 \times 10^{-10}$~g/cm$^3$,
in accordance with estimates of the initial mass planetary nebula \citep{Hayashi1981} at an Earth orbit,
the outer boundary density thus drops to a value of $\rho = 5 \times 10^{-25}$~g/cm$^3$.

Technically, these very low densities are in a range
where the fluid approximation of hydrodynamics becomes invalid.
In our calculations, we neglected this effect to represent, in the continuum description, the scenario of
$\rho \rightarrow 0$ (and thus $P \rightarrow 0$) at the outer margin of the planetary atmosphere.
Moreover, in reality the very thin
outer region of the atmosphere is dominated by stellar wind and stellar irradiation.
Consequently, the gas is ionized to a significant degree and
a plasma can exhibit collective phenomena arising out of mutual
interactions of many charged particles.
This is usually described by the so-called dimensionless plasma parameter $\Lambda$,
which basically characterizes the number of particles within a Debye sphere,
i.e., $\Lambda = n \; 4 \pi \lambda_{\rm D}^3$, where $n$ is the number density of electrons and
$\lambda_{\rm D}$ is the Debye length.
The larger this plasma parameter
is (typically $\Lambda \simeq 10^6$ in the interplanetary medium), the more collective
interactions are taking place because of the decreasing effective shielding volume.
Hence, a situation is established that still allows for a
hydrodynamical description of the fluid \citep[e.g.,][]{Choudhuri1998}.

In terms of astrophysical evolution of planetary systems, the evaporation of disks is a quite fast process.
Statistics of transitional disks indicate a characteristic transition timescale of about $(0.1-1) \times 10^6$ years \citep{Muzerolle2010}.
As we are interested in the dynamic reaction of the atmosphere,
we have to implement this density change on timescales smaller than
the dynamical timescale of the atmosphere,
independent of the realistic evaporation timescale of actual disks.
In Sect.~\ref{SECT_dynamic_results}, we show that for our atmospheric models the time to adjust to
changes in the outer boundary density ranges between $\sim 10^{2}$ and $\sim 10^{13}$ years.
For consistency and comparability, the evaporation of the disk is included in all dynamical simulations
on a common timescale shorter than this range, in particular, over a period in time of $10^{9}$ s ($\simeq 32$ yr).
The temperature at the boundary condition is kept fixed at 200~K,
which is about the radiation equilibrium temperature at an Earth-like orbit at 1~AU
around a young solar-type star throughout the whole simulation run.

%%%%%%%%%%%%%%%%%%%%%%%%%%%%%%%%%%%%%%%%%%%%%%%%%%%%%%%%%%%%%%%%%%%%%%%%%%%%%%%%%%%%%%%%%%%%%%%%%%%%
\section{Stationary models}
\label{SECT_stationary}
%%%%%%%%%%%%%%%%%%%%%%%%%%%%%%%%%%%%%%%%%%%%%%%%%%%%%%%%%%%%%%%%%%%%%%%%%%%%%%%%%%%%%%%%%%%%%%%%%%%%

\subsection{Method}
The stationary models are computed from the hydrostatic structure equations
\begin{equation}
\pp{m}{r} = 4 \pi r^2 \rho \, ,
\end{equation}
\begin{equation}
\pp{P}{r} = -\frac{G m}{r^2} \rho \; ,
\end{equation}
and
\begin{equation}
\pp{T}{r} = \pp{P}{r} \; \cdot \;\frac{T}{P} \nabla \: .
\end{equation}
The meaning of the symbols we used is complied in Tab.~\ref{TAB_meaning}
These equations are solved using the initial model integrator of the
TAPIR-Code ({\bf T}he {\bf a}da{\bf p}tive, {\bf i}mplicit {\bf R}HD-Code).
The TAPIR-Code provides a framework for the implicit solution of the equations of radiation
hydrodynamics on adaptive grids and previously has been applied to
spherical flows \citep{Dorfi2006},
2-dimensional flows \citep{Stoekl2007},
convection modeling in Cepheids \citep{Stoekl2008},
and planetary atmospheres \citep{Erkaev2013b, Lammer2014}.

The logarithmic temperature gradient $\nabla$ is either determined from radiative transport
\begin{equation}
\nabla_{\rm rad} = \frac{3 \kappa_{\rm R} L_{\rm pl} P }{64 \pi \sigma G m T^4}
\end{equation}
or, in case of convective instability, calculated from the turbulent convection model yielding a
temperature gradient $\nabla_{\rm conv}$, usually close to $\nabla_{\rm ad}$.
In case of the integration of hydrostatic models, we use the stationary limit of
the time-dependent formulation of the turbulent convective energy equation given below in Eq.~\ref{EQ_trubeneeq}.

\begin{table}
\caption{Meaning of Symbols.}
\label{TAB_meaning}
{ \scriptsize
\begin{tabular}{cl}
\hline
\noalign{\smallskip}
Symbol & Meaning \\
\noalign{\smallskip}
\hline
$r$               & Radius from the center of the planet \\
$t$               & Time \\
$m$               & Integrated mass (core mass + atmospheric mass)\\
$P$               & Gas pressure \\
$T$               & Gas temperature \\
$\rho$            & Gas density \\
$e$               & Specific gas internal energy \\
$\vec{u}$         & Gas velocity \\
$u'$              & Convective velocity perturbation \\
$J$               & Zeroth moment of radiation intensity ($\propto$ radiation energy)\\
$\vec{H}$         & First moment of radiation intensity ($\propto$ radiation flux)\\
$\tens{K}$        & Second moment of radiation intensity ($\propto$ radiation pressure) \\
$\Phi$            & Gravitational potential \\
$\sigma$          & Stefan–Boltzmann constant \\
$G$               & Gravitational constant \\                           % doppelt erlaeutert
$\mathcal{R}$     & Specific gas constant \\                            % doppelt erlaeutert
$c_P$             & Specific heat at constant pressure \\
$\gamma$          & Adiabatic index \\
$c_{\rm s}$       & Adiabatic sound speed \\
%$F_{\rm conv}$    & Convective flux \\                                 % doppelt erlaeutert
%$S$               & Source function of radiation \\                    % doppelt erlaeutert
%\tens{Q}          & Viscous pressure tensor \\                         % doppelt erlaeutert
$\kappa_{\rm R}$  & Total (gas \& dust) opacity, Rosseland-mean \\      % doppelt erlaeutert
$\kappa_{\rm P}$  & Total (gas \& dust) opacity, Planck-mean \\         % doppelt erlaeutert
\noalign{\vspace{1pt}}
$\nabla = \pp{\ln{T}}{\ln{P}}$ & Logarithmic temperature gradient \\    % doppelt erlaeutert
\noalign{\vspace{1pt}}
$L_{\rm pl}$      & Planetary luminosity \\                             % doppelt erlaeutert
$M_{\rm pl}$      & Mass of the planetary core \\                       % doppelt erlaeutert
$M_{\star}$       & Mass of the host star \\                            % doppelt erlaeutert
\noalign{\smallskip}
\hline
\end{tabular}
}
\end{table}

As noted in the introduction,
spherical symmetry is a good assumption for $r \ll R_{\rm Hill}$ and breaks down completely at $r > R_{\rm Hill}$,
therefore, the natural point to implement the boundary conditions is the Hill radius.
Some previous studies \citep[e.g.,][]{Ikoma2006,Rafikov2006,Ikoma2012} %auch Ikoma, Emori & Nakazawa 2001
used the minimum of Hill radius and Bondi radius
(where the latter typically is about one order of magnitude smaller)
for the placement of the outer boundary.
The Bondi radius relates the the gravitational potential to the enthalpy of the gas and thus provides
an indication to which extent the atmosphere is gravitationally bound to the planet
\begin{equation}
R_{\rm Bondi} = \frac{G\;M_{\rm pl} }{c_{\rm s}^2} \; ,
\end{equation}
where $c_{\rm s}$ is the speed of sound in the nebula.
The Bondi radius provides a good estimate for the point where the atmospheric density structure
attaches to the surrounding nebula gas.
There is, however, no reason not to extend hydrostatic models beyond the Bondi radius into the nebula environment
up to the ultimate limit of spherical modeling at the Hill radius \citep{Mizuno1978}.
In view of the dynamic calculations in Sect.~\ref{SECT_dynamic},
where it is instrumental to include a domain beyond the immediate planetary envelope structure,
we placed the outer boundary of our models on the Hill radius throughout.
For the calculation of the Hill radius, we adopted $a= 1 \rm{AU}$ and $M_{\star} = 1 M_{\odot}$ as standard parameters.

The accretion rate of planetesimals, and in consequence the planetary luminosity,
is unfortunately a highly uncertain parameter.
Almost the only indication comes from the observed disk lifetime of some $10^{6}$ years
and the fact that the planetary cores must have formed within that period of time.
Accordingly, we adopted a mass accretion rate $\dot{M}_{\rm acc}/M_{\rm pl}$ of $10^{-7}{\rm yr}^{-1}$ as the standard value.

In addition to accretion, the models also allow for radiogenic heat production,
which is included proportional to the core mass
using a reference value of $10^{21}$ erg s$^{-1}$ $M_{\rm \oplus}^{-1}$ for the young Earth \citep{Stacey2008}.
This contribution, however, never becomes relevant as the accretion luminosity is
more than two orders of magnitude larger for all models calculated.

For the atmospheric gas, we assumed a homogeneous chemical composition with solar abundances. The data for
gas opacity is taken from \citet{Freedman2008} and for the equation of state we adopted the results of \citet{Saumon1995}.
Dust is included in the model with a fixed dust grain depletion factor $f$
using dust opacity coefficients from \citet{Semenov2003}.
The dust depletion factor $f$ specifies the amount of dust in the atmosphere relative to the
local dust condensation/evaporation equilibrium.
As dust opacities dominate over gas opacities by many magnitudes in large parts of the atmosphere,
$f$ in effect scales the overall opaqueness of the atmosphere.
To allow for dust depletion processes such as rain-out in the atmosphere and dust segregation
within the circumstellar disk structure, $f$ is usually set to values $f<1$ and we
adopted a standard value of $f = 0.01$.

%%%%%%%%%%%%%%%%%%%%%%%%%%%%%%%%%%%%%%%%%%%%%%%%%%%%%%%%%%%%%%%%%%%%%%%%%%%%%%%%%%%%%%%%%%%%%%%%%%%%
\subsection{Results}
\label{SECT_stationary_results}

\begin{figure}
\includegraphics[height=1.\columnwidth,angle=-90]{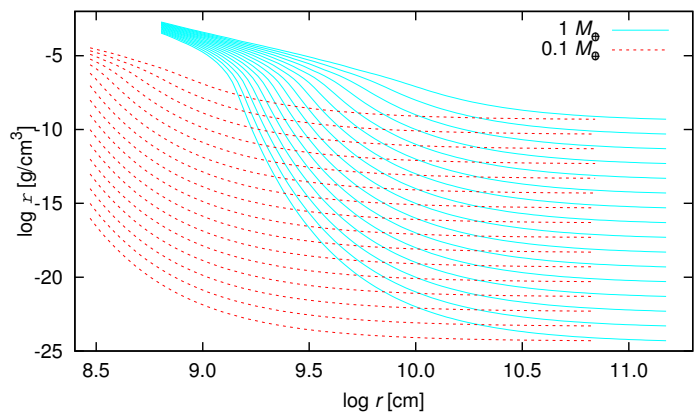}
\caption{
Density as a function of radius in protoplanetary atmospheres around cores of
0.1~$M_{\rm \oplus}$ and 1~$M_{\rm \oplus}$
and for nebula gas densities varying from $5 \times 10^{-10}$~g/cm$^3$ to $5 \times 10^{-25}$~g/cm$^3$ with steps of one magnitude.
The inner and outer boundaries are defined by the radius of the solid core and by the Hill radius, respectively.
}
\label{FIG_structure-rhoOBC}
\end{figure}

Figure~\ref{FIG_structure-rhoOBC} illustrates the dependence of the hydrostatic atmosphere structures on density variations
at the outer boundary for cores of 0.1~$M_{\rm \oplus}$ and 1~$M_{\rm \oplus}$. The very different behavior in the two model sequences is striking:
whereas the atmosphere around the Mars-mass core almost completely escapes with the removal of the surrounding nebula gas,
the structure of the atmosphere around the Earth-mass only changes gradually and substantial parts of the atmosphere remain gravitationally bound.
This confirms the results of \citet{Ikoma2006}, who found an almost linear relation between
atmospheric mass and nebula density for a Mars-mass core
in contrast to a much less dependent atmospheric mass for an Earth-mass planet.
In terms of absolute numbers, our results are in agreement with those from \citet{Ikoma2006}.
The main physical difference between this study and \citet{Ikoma2006} is the data for dust and gas opacities.
Other numerical differences, such as the treatment of convective transport and
the placement of the outer boundary condition, are less important.

\begin{figure}
\includegraphics[height=1.\columnwidth,angle=-90]{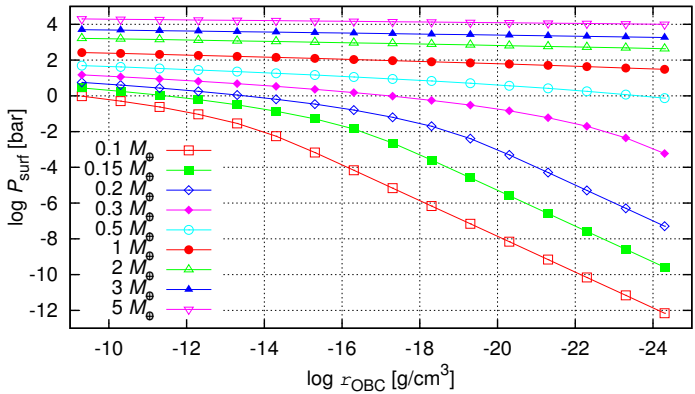}
\caption{Gas pressure at the bottom of the atmosphere on the surface of solid cores
with masses between 0.1~$M_{\rm \oplus}$ and 5~$M_{\rm \oplus}$ as a function of the nebula gas density $\rho_{\rm OBC}$.
% The response of the atmospheres on the changing outer boundary density varies with core mass between almost
% complete escape and comparatively small readjustments.
%  model                      P_start/P_end
%  M0.1_MAcc1E-8_dust0.01     1.382E+12
%  M0.15_MAcc1.5E-8_dust0.01  1.067E+10
%  M0.2_MAcc2E-8_dust0.01     1.098E+08
%  M0.5_MAcc5E-8_dust0.01     6.618E+01
%  M1_MAcc1E-7_dust0.01       8.778E+00
%  M2_MAcc2E-7_dust0.01       3.819E+00
%  M3_MAcc3E-7_dust0.01       2.731E+00
%  M5_MAcc5E-7_dust0.01       1.979E+00
}
\label{FIG_P-rhoOBC}
\end{figure}

Figure~\ref{FIG_P-rhoOBC} illustrates the disparity in atmospheric behavior
for a broader range of planetary core masses.
As a measure for the amount of atmosphere, we use in Fig.~\ref{FIG_P-rhoOBC}
the gas pressure on the surface of the solid core.
This avoids the ambiguity of evaluating the mass of the atmosphere, separating the gravitationally accumulated gas from the
background density structure of the circumstellar disk.
Judging by the surface pressure, we see, first of all, that the amount of atmosphere increases with the mass of the planetary core.
The core with 5~$M_{\rm \oplus}$ having at least 4 orders of magnitude more atmosphere than the 0.1~$M_{\rm \oplus}$ core.

In Fig.~\ref{FIG_P-rhoOBC}, the findings of Fig.~\ref{FIG_structure-rhoOBC} are now placed in a broader scheme.
Dependent on the mass of the planetary core,
the reaction of the atmospheres varies between almost complete escape and comparatively small readjustments.
For the core with 0.1~$M_{\rm \oplus}$, the surface pressure reduces by more than 12 orders of magnitude,
for the 1~$M_{\rm \oplus}$ core by a factor of about 9,
and by a factor smaller than 2 for the 5~$M_{\rm \oplus}$ core.
In this simple, stationary analysis, planetary cores with masses smaller than about 0.5~$M_{\rm \oplus}$ seem
unable to retain a substantial part of the nebula-accumulated atmosphere once the nebula disappears.
Similar sequences of atmosphere models for a 1~$M_{\rm \oplus}$ core have been presented by \citet{Nakazawa1985} and \citet{Ikoma2006}.

Atmospheres around more massive cores are less sensitive to variations of the outer boundary density.
This comparative insensitiveness on the outer boundary conditions
is an effect which, in similar form, has also been observed
in earlier studies of planetary atmospheres with radiative outer envelopes
\citep[e.g.,][]{Stevenson1982, Wuchterl1993, Rafikov2006}.

The general outline of the results does not depend on the particular choice of physical and numerical parameters,
such as host star and disk properties, planetary orbit, dust opacities, or accretion rate of planetesimals.
Different parameters basically shift the scenario in the $M_{\rm pl}$ -- $L_{\rm pl}$ plane
but leave it otherwise unchanged qualitatively.

To understand the structures and properties of the atmospheres shown in Fig.~\ref{FIG_structure-rhoOBC},
we first note that
self-gravity is not important in these low-mass atmospheres.
The atmospheres therefore can be considered to
reside in
a point-source gravitational potential in good approximation.
In this picture, the atmospheric structure is independent from the particulars of the solid core and
the surface radius of the solid core essentially becomes an arbitrary radius where the atmospheric structure is truncated.

Another important observation is that,
with respect to stratification, the atmospheres
can be separated into three distinctive parts:
(1) the inner optical thick region where convective energy transport takes place and
the adiabatic stratification of which
can be approximated by a polytropic structure with a polytropic index of $n \simeq 3.3$;
(2) an intermediate radiative region where the stratification is determined by radiative transport;
and
(3) the outer part of the atmosphere that is optically thin and therefore
accurately described by an isothermal structure (i.e., $n = \infty$).

%          | # FhG | gamma |  n
%----------+-------+-------+------
% 1-atomig |   3   |  5/3  | 1.5
% 2-atomig |   5   |  7/5  | 2.5
% 3-atomig |   6   |  4/3  | 3
%
%   n     = 1/(gamma-1)
%   gamma = (n+1)/n

The total opacity in the intermediate radiative region is usually dominated by the dust opacity, which
varies in a series of dust layers, corresponding to the formation and evaporation of dust species.
This leads, on the one hand, to a substantial, nonzero temperature gradient, varying according to the local opacity.
On the other hand, the maximum value of the temperature gradient in the high-opacity regions,
is limited at the adiabatic value by the onset of convective transport.
On average, the temperature gradient in the intermediate radiative region is therefore,
somewhat smaller than, but not far off the adiabatic temperature gradient.
As compared to the outer, optically thin part of the envelope, the stratification in the
intermediate radiative part is quite similar to the inner convective part and in the temperature profiles, shown in
Fig.~\ref{FIG_M01dyn} and Fig.~\ref{FIG_M2dyn}, the boundaries of the convective regions are essentially indiscernible.
It is therefore reasonable to combine both the inner convective region and the optical thick radiative region above
in one approximated polytropic structure with a common polytropic index.
That way, the planetary envelope is dissected into a two-part analytical model
where the inner optical thick part of the atmosphere is described as a polytropic structure
\citep[e.g.,][]{Eddington1926}
\begin{equation}
\label{EQ_polytrope}
\rho(r) = \left( C_1 + C_2 \frac{G M_{\rm pl}}{r}\right)^n \; ,
\end{equation}
while the density of the optical thin and isothermal outer part can be written as
\begin{equation}
\label{EQ_isotherm}
\rho(r) = C_3 \exp{ \left( \frac{1}{\mathcal{R}T} \frac{G M_{\rm pl}}{r} \right) } \; .
\end{equation}
Here, $n$ is the polytropic index of the inner part, $T$ the gas temperature of the isothermal outer part,
and $C_1$, $C_2$, and $C_3$ are constants.

\begin{figure}
\includegraphics[height=1.\columnwidth,angle=-90]{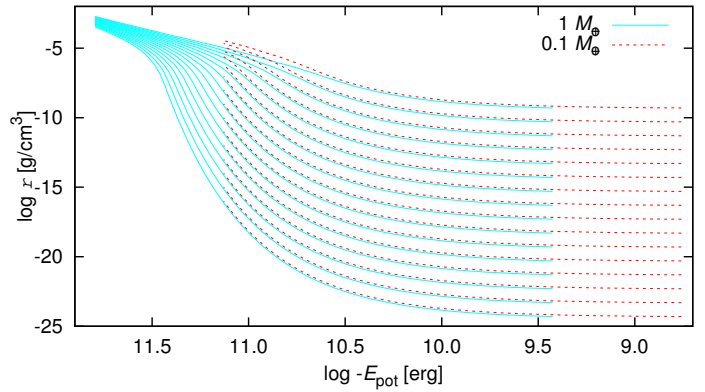}
\caption{
Density as a function of the gravitational potential
for the same atmospheric models as in Fig.~\ref{FIG_structure-rhoOBC}
for cores masses of 0.1~$M_{\rm \oplus}$ and 1~$M_{\rm \oplus}$ and nebula gas densities between
$5 \times 10^{-10}$~g/cm$^3$ to $5 \times 10^{-25}$~g/cm$^3$.
}
\label{FIG_structure_pot-rhoOBC}
\end{figure}

In this description, the density is solely a function of the gravitational potential.
Figure~\ref{FIG_structure_pot-rhoOBC}, which
plots the same sequences of atmospheric structures
as in Fig.~\ref{FIG_structure-rhoOBC} against the gravitational potential,
confirms that this also holds for the numerical models in good approximation.
Even though important physical ingredients of the numerical models, such as
accretion luminosity, convection model, and realistic data for the equation of state and opacities,
do not scale in a simple manner,
the similarity between the two sets of solutions is very clearly apparent.

From this perspective, it appears that the atmospheric structures around
cores of different mass are essentially homologous,
but shifted in the radius coordinate by the different core mass and truncated at different
relative positions according to the radius of the respective core.

The different dependency of the atmospheric structures on the outer boundary density can also be explained from the
simple two-part analytical model:
starting at a outer boundary with
a density $\rho_{\rm OBC}$,
a temperature $T_{\rm OBC}$, and
a radius $r_{\rm OBC}$,
we can calculate the constant $C_3$ in Eq.~\ref{EQ_isotherm} and
obtain for the density profile in the outer isothermal part
\begin{equation}
\label{EQ_outerpart}
\rho = \rho_{\rm OBC} \exp{ \left( \frac{1}{\mathcal{R}T_{\rm OBC}} \left[ \frac{G M_{\rm pl}}{r} - \frac{G M_{\rm pl}}{r_{\rm OBC}} \right] \right) } \; .
\end{equation}
With increasing density toward the surface of the planet,
the atmosphere gets more opaque and at a density $\rho_{\rm Trans}$
the isothermal structure changes to a polytropic stratification.
In the context of the two-part analytical model, $\rho_{\rm Trans}$ has to be specified as a parameter.
The effect of $\rho_{\rm Trans}$ is that it
scales the density in the deep atmosphere and thus the total mass of the atmosphere.

The gravitational potential at this transition point $(G M_{\rm pl}/r_{\rm Trans})$ is a function of the outer boundary density $\rho_{\rm OBC}$.
The temperature at the transition point $T_{\rm Trans}$ is that of the isothermal part, i.e., $T_{\rm Trans} = T_{\rm OBC}$.
From these interface values, $r_{\rm Trans}$, $\rho_{\rm Trans}$, and $T_{\rm Trans}$
and integrating the pressure in a polytropic structure, we can calculate
$C_1$ and $C_2$ in Eq.~\ref{EQ_polytrope} and rewrite the density profile for the inner polytropic part as
\begin{equation}
\label{EQ_innerpart}
\rho = \rho_{\rm Trans} \left( 1 + \frac{1}{\mathcal{R}T_{\rm OBC}} \frac{1}{n+1} \left[ \frac{G M_{\rm pl}}{r} - \frac{G M_{\rm pl}}{r_{\rm Trans}} \right] \right)^n \; .
\end{equation}
The only point where the outer boundary density $\rho_{\rm OBC}$ enters this equation is via $r_{\rm Trans}$.
For large gravitational potentials $\frac{G M}{r} \gg \frac{G M}{r_{\rm Trans}}$ (i.e., $r \ll r_{\rm Trans}$) the density converges to
\begin{equation}
\label{EQ_veryinnerpart}
\rho = \rho_{\rm Trans} \left( 1 + \frac{1}{\mathcal{R}T_{\rm OBC}} \frac{1}{n+1} \frac{G M_{\rm pl}}{r} \right)^n \; ,
\end{equation}
independent from the density at the outer boundary.
Therefore, the more massive a planetary core is, the less dependent is its deep atmosphere on the outer boundary density.
This also explains, as the mass of a planetary atmosphere is concentrated at the bottom of the atmosphere, the decreasing dependency of
the atmosphere on the outer boundary density for more massive planets as illustrated in Fig.~\ref{FIG_P-rhoOBC}.

\begin{figure}
\includegraphics[height=1.\columnwidth,angle=-90]{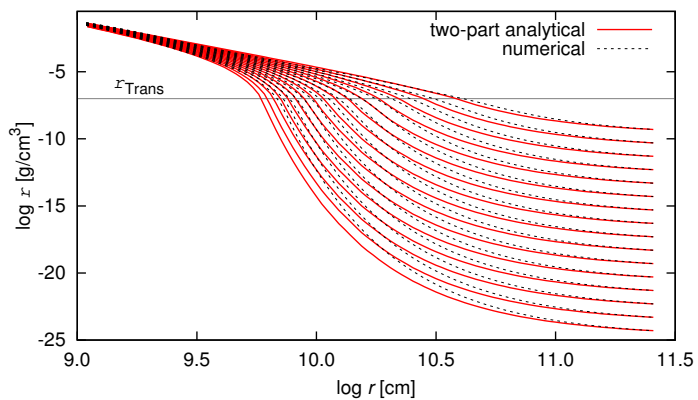}
\caption{Density as a function of radius for an atmosphere around a 5~$M_{\rm \oplus}$ core, as computed
from the detailed numerical description (TAPIR)
and from the two-part analytical model (see text).
}
\label{FIG_structure-2part}
\end{figure}

Figure \ref{FIG_structure-2part} compares the two-part analytical model for a polytropic index $n=3.3$
and a transition density of $\rho_{\rm Trans} = 10^{-7}$~g/cm$^3$ with the numerical results.
For this comparison, we selected the most massive planet with 5~$M_{\rm \oplus}$ from our sequence to illustrate the converging
density profiles in the deep atmosphere.
The transition points between the outer isothermal and the inner polytropic part can be read from Fig.~\ref{FIG_structure-2part}
from the intersections of the density profiles with the horizontal line indicating the transition density.

The very simple analytical model reproduces the numerical models surprisingly well.
Moreover, the two constitutional parameters of the analytical description are closely constrained;
$n$ from the atmospheric composition and
$\rho_{\rm Trans}$ scales the amount of atmosphere, in a similar way as the luminosity does for the numerical models.

%%%%%%%%%%%%%%%%%%%%%%%%%%%%%%%%%%%%%%%%%%%%%%%%%%%%%%%%%%%%%%%%%%%%%%%%%%%%%%%%%%%%%%%%%%%%%%%%%%%%
\section{Dynamic outflow simulations}
\label{SECT_dynamic}
%%%%%%%%%%%%%%%%%%%%%%%%%%%%%%%%%%%%%%%%%%%%%%%%%%%%%%%%%%%%%%%%%%%%%%%%%%%%%%%%%%%%%%%%%%%%%%%%%%%%

\subsection{Method}
\label{SECT_dynamic_method}

For the calculation of the time-dependent atmosphere models,
the equations of radiation hydrodynamics are solved
by the TAPIR-Code using an implicit time integration scheme.
The discretization of the physical equations is based on finite volumes on a staggered mesh
and advective flows are calculated using the second order van Leer flux limiter.

Our system of equations consists of the equation of continuity
\begin{equation}
\label{rhdeq.eq1}
\frac{\partial}{\partial t} \rho + \vec \nabla \cdot ( \vec u \, \rho ) = 0 \; ,
\end{equation}
\noindent the equation of motion
\begin{equation}
\label{rhdeq.eq2}
\pp{}{t} ( \rho \vec u )
   + \vec \nabla \cdot ( \vec u \, \rho \vec u )
   + \vec \nabla P
   + \rho\vec{\nabla}\phi
   - \frac{4 \pi}{c} \kappa_{\rm R} \rho \vec{H}
   + \vec \nabla \cdot \tens{Q} = 0 \; ,
\end{equation}
\noindent the equation of internal energy
\begin{multline}
\label{rhdeq.eq3}
\pp{}{t} ( \rho e )
  + \vec \nabla \cdot ( \vec u \, \rho e )
  + P \, \vec \nabla \cdot \vec u
  - 4 \pi \kappa_{\rm P} \rho ( J - S ) + \\
  + \tens{Q} : \vec \nabla \vec u
  + \vec \nabla \cdot F_{\rm conv} = 0 \; ,
\end{multline}
\noindent the Poisson equation
\begin{equation}
\label{rhdeq.eq4}
\Delta \phi = 4 \pi G \rho \; ,
\end{equation}
\noindent the radiation energy equation (zeroth moment of radiation intensity)
\begin{equation}
\label{rhdeq.eq5}
\pp{}{t} J
+ \vec \nabla \cdot ( \vec u \, J )
+ c \, \vec \nabla \cdot \vec H
+ \tens{K} : \vec \nabla \vec u
+ c\; \kappa_{\rm P} \rho ( J - S ) = 0 \; ,
\end{equation}
\noindent and the radiation flux equation (first moment of radiation intensity)
\begin{equation}
\label{rhdeq.eq6}
\pp{}{t} \vec H
+ \vec \nabla \cdot ( \vec u \, \vec H )
+ c \, \vec \nabla \cdot \tens{K}
+ \vec H \cdot \vec \nabla \vec u
+ c \; \kappa_{\rm R} \rho \vec H = 0 \; .
\end{equation}
For the meaning of the symbols see Tab.~\ref{TAB_meaning}.
In particular, $S$ is the source function of radiation with $S = \frac{\sigma}{\pi} T^4$ and
$\tens{Q}$ is the viscous pressure tensor defined as
\begin{equation}
 \tens{Q} = -q_{\rm lin} \; r \rho c_{\rm s}
            \biggl[ \frac{1}{2}\left( \vec \nabla \vec u + (\vec \nabla \vec u)^{\rm T} \right) - \frac{1}{3} \, \vec \nabla \cdot \vec u \biggr]
\end{equation}
and scaled by the viscosity parameter $q_{\rm lin}$.
The equation of state enters the system of radiation hydrodynamics equations via
the gas temperature $T$ and the gas pressure $P$.
Gas and dust opacities appear both in the form of the Rosseland-mean $\kappa_{\rm R}$ and Planck-mean $\kappa_{\rm P}$.

The closure of the moment description of the radiation field $(J,\vec H,\tens{K})$ requires the
specification of the Eddington factor $\tens{f}_{\rm edd} = \tens{K} / J$
where we use $f_{\rm edd} = \frac{1}{3}$, i.e., the Eddington approximation, throughout.

Convective energy transport is included in TAPIR through an equation for the turbulent convective energy
following concepts of \citet{Kuhfuss1987} along the lines of \citet{Wuchterl1998}.
Ignoring nonessential additions
(turbulent pressure, convective viscosity, diffusion of turbulent convective energy, radiation pressure in the source term),
the model takes the form of
\begin{multline}
\label{EQ_trubeneeq}
\pp{}{t} \left( \rho \tfrac{1}{2} {u'}^2 \right) + \vec \nabla \cdot \left( \vec u \, \rho \tfrac{1}{2} {u'}^2 \right)
% + \tens{Q}_{turb} : \vec \nabla \vec u
% + P_{turb} \vec \nabla \cdot \vec u
% - \alpha_{di\!f\!f} \vec \nabla \cdot \left( \sqrt{\frac{1}{3} } u' \rho \Lambda \; \pp{\frac{1}{2} {u'}^2}{r} \right) \nonumber \\
 + \rho \frac{T}{P} \nabla_{\rm \! ad} \; \vec \nabla P \cdot \langle s' \vec u' \rangle + \\
 + \alpha_{\rm diss} \, \rho \frac{ \frac{1}{2} {u'}^3 }{\Lambda}
 + \frac{4\sigma T^3 \alpha_{\rm rad}^2}{c_P \kappa_{\rm R} \rho \Lambda^2 } \tfrac{1}{2} {u'}^2 = 0 \; .
\end{multline}
The correlation between entropy perturbation and velocity perturbation is provided by the diffusion ansatz
\begin{equation}
\left \langle s' \vec u' \right \rangle \simeq - \alpha_{\rm s} \Lambda \sqrt{\frac{1}{3}} \; u' \displaystyle{ \pp{s}{r} } \; ,
\end{equation}
where the entropy gradient is evaluated as
\begin{equation}
\pp{s}{r} = \frac{1}{T} \pp{e}{r} - \frac{P}{\rho^2 T} \pp{\rho}{r}
\end{equation}
and the convective flux is computed from
\begin{equation}
F_{\rm conv} = \rho T \left \langle s' \vec u' \right \rangle \; .
\end{equation}
Table~\ref{TAB_convparas} lists the standard parameters adopted for all calculations, which have been
adjusted so as to reproduce the solutions of the mixing length theory
\citep[in the formulation of][]{KippenhahnWeigert1990} in the stationary limit.

\begin{table}
\caption{Standard parameters for the turbulent convection model.}
\label{TAB_convparas}
{ \scriptsize
\begin{tabular}{ccl}
\hline
\noalign{\smallskip}
Parameter & Adopted values & Physical meaning \\
\noalign{\smallskip}
\hline
\noalign{\smallskip}
$\Lambda$            & $\alpha_{\rm ML}$ = 2 & Mixing length, computed from $\alpha_{\rm ML}$\\
\noalign{\smallskip}
$\alpha_{\rm s}$     & 0.866                 & Diffusion parameter for entropy perturbation \\
\noalign{\smallskip}
$\alpha_{\rm diss}$  & 4                     & Dissipation of turbulent energy \\
\noalign{\smallskip}
$\alpha_{\rm rad}$   & 9                     & Radiative losses \\
\noalign{\smallskip}
\hline
\end{tabular}
}
\end{table}

Finally, the system of physical equation can be supplemented with an additional grid equation \citep{Dorfi1987},
which is solved together with the system of physical equations.
The grid equation governs the motion of a fixed number of grid points (usually 500 points)
and continuously redistributes them to allow for an
optimal resolution of the physical gradients within the evolving solution.

The set of equations is solved with an
implicit scheme, which repeatedly computes
corrections to the set of variables from the
inversion of the Jacobi-matrix until a relative accuracy of $10^{-5}$ is achieved.
Time step control essentially aims for
computational efficiency by balancing time step size with
computational effort, i.e., the number of iterations required for a single step.
The implicit time integration method allows for large time steps without
being restricted by the Courant-Friedrichs-Lewy condition
and therefore facilitates the efficient calculation of both fast dynamical processes
as well as of long-term evolutionary processes.
The simulations were terminated at an age of $10^{16}$ s as all runs are in a stationary state by then.
To cover this period in time typically requires a few thousand time steps.

The physical scenario of the time-dependent simulations is the same as for the stationary models:
i.e., the computational domain spans from the surface of the planetary core up to the Hill radius
where the outer boundary conditions are applied
, in particular $\rho = 5 \times 10^{-10}$~g/cm$^3$, $T = 200$~K, and $J=S$.
The Hill radius is based on an orbit with $a= 1 \rm{AU}$ and a host star with $M_{\star} = 1 M_{\odot}$.
The outer boundary is open for fluid flow and radiation and
the solution of the equation of motion and radiation flux equation at the outer boundary
gives the mass and energy exchange with the disk nebula environment.

The dust depletion factor is fixed at $f = 0.01$, and for the planetary accretion rate of planetesimals
we used a constant mass accretion rate of $\dot{M}_{\rm acc}/M_{\rm pl} = 10^{-7}{\rm yr}^{-1}$.
Here, the accretion rate of planetesimals is considered solely as a motivation for the planetary luminosity,
the mass of the solid core is not increased in consequence of accretion (and thus neither is the Hill radius).
This inconsistency is unavoidable when simulating over the very long periods in time necessary to
reach stationary solutions while using the simple correlation between $\dot{M}_{\rm acc}$ and $L_{\rm pl}$.
Whereas in the stationary models, the adopted planetary luminosity is constant throughout the atmospheric structure,
we assume, in the time-dependent models, that it originates at the bottom of the atmosphere on the surface of the solid core.

As for the stationary models, we
assumed a chemically homogeneous medium with solar composition and
used gas opacities from \citet{Freedman2008}, dust opacities from \citet{Semenov2003}, and the
\citet{Saumon1995} equation of state.

%%%%%%%%%%%%%%%%%%%%%%%%%%%%%%%%%%%%%%%%%%%%%%%%%%%%%%%%%%%%%%%%%%%%%%%%%%%%%%%%%%%%%%%%%%%%%%%%%%%%
\subsection{Results}
\label{SECT_dynamic_results}

\begin{figure}
\includegraphics[height=1.\columnwidth,angle=-90]{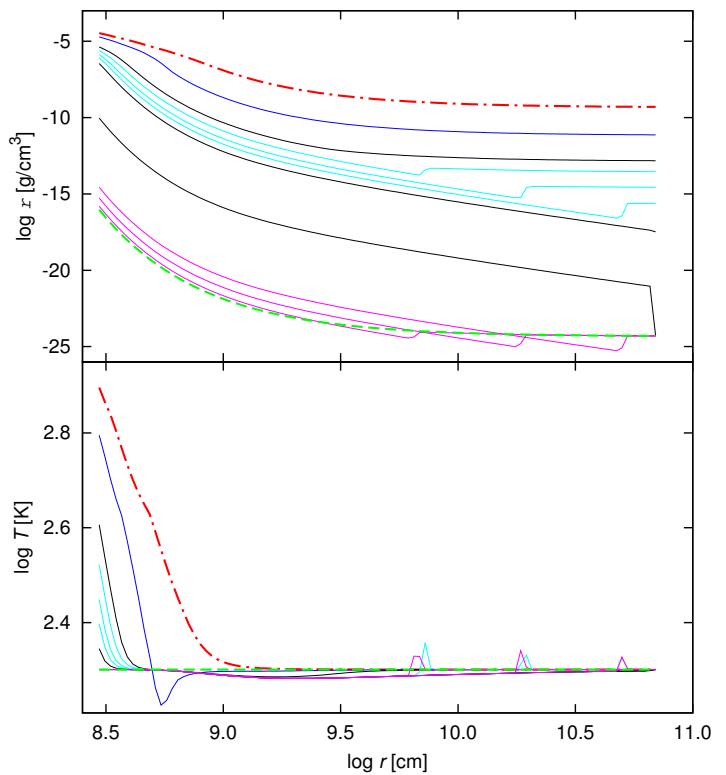}
\caption{
Density and temperature as a function of radius for the atmosphere around an 0.1~$M_{\rm \oplus}$ core
at several snapshots during the dynamic outflow caused by a drop of the
density at the outer boundary from $5 \times 10^{-10}$~g/cm$^3$ to $5 \times 10^{-25}$~g/cm$^3$
over $\simeq 32$ years.
The dash-dotted lines are the initial model, and after the dynamic outflow phase the configuration again settles
in a hydrostatic configuration indicated by dashed lines.
The times of the snapshot are, from top to bottom, in years:
0 (dash-dotted line),
3.9,
7.5,
9,
11,
13,
17,
64,
141,
148,
154,
$\infty$ (dashed line).
In the color version of this figure, some lines are highlighted in color
to simplify the identification of equivalent snapshots in the two panels.
}
\label{FIG_M01dyn}
\end{figure}

\begin{figure}
\includegraphics[height=1.\columnwidth,angle=-90]{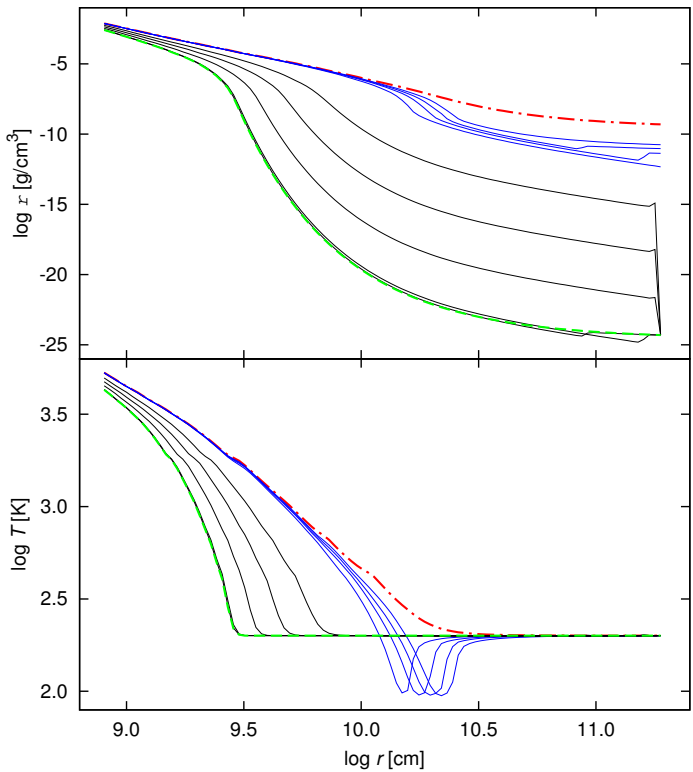}
\caption{Same as Fig.~\ref{FIG_M01dyn}, except for a core with 2~$M_{\rm \oplus}$.
The times of the snapshot are, from top to bottom, in years:
0 (dash-dotted line),
3.1,
3.7,
4.4,
6.4,
$2.3 \times 10^{3}$,
$1.3 \times 10^{6}$,
$1.6 \times 10^{9}$,
$1.9 \times 10^{12}$,
$2.9 \times 10^{12}$,
$\infty$ (dashed line).
In the color version of this figure, some lines are highlighted in color
to simplify the identification of equivalent snapshots in the two panels.
}
\label{FIG_M2dyn}
\end{figure}

The physical scenario of the time-dependent simulations is again the dynamic outflow process
as described in Sect.~\ref{SECT_scenario}.
The evolving structure of the atmospheres is shown in two examples in
Fig.~\ref{FIG_M01dyn} and Fig.~\ref{FIG_M2dyn} for
planetary cores with 0.1~$M_{\rm \oplus}$ and 2~$M_{\rm \oplus}$, respectively.

Apart from the imprints of dynamical processes, discussed below, the temperature profiles (lower panels)
are characterized by an isothermal outer part and an essentially continuous temperature profile
in the inner optical thick radiative and convective regions.

In the density profiles (upper panels of Fig.~\ref{FIG_M01dyn} and Fig.~\ref{FIG_M2dyn}),
two different types of stratification are discernible:
quasi-hydrostatic and outflow-dominated.
In the former case, the stratification is still essentially in hydrostatic equilibrium,
i.e., the density gradient is similar to that of the corresponding hydrostatic model.
In the latter case, the structure is no longer in hydrostatic balance but dominated by the outflow,
which results in a density gradient according to the solution of an isothermal wind \citep{Lamers1999}.
In particular in the outer part of the upper panel of Fig.~\ref{FIG_M01dyn}, the two different gradients are clearly apparent.

In general, the interfaces between these two solutions would be discontinuous shocks,
but here they are broadened by numerical dissipation and appear as bumps of finite width.
The magnitude of these bumps in the density profile is characterized by
a constant total pressure (composed of gas and dynamic pressure),
i.e., $P_{\rm tot} = P_{\rm gas} + P_{\rm dyn} = P_{\rm gas} + \tfrac{1}{2} \rho u^2 \simeq \text{const.}$.
The energy dissipation in these shocks
is a source term to the internal energy and thus may produce spikes in the temperature profile,
depending on the local efficiency of radiative transport.

In Fig.~\ref{FIG_M01dyn} and Fig.~\ref{FIG_M2dyn}, the simulation starts
from an initial hydrostatic model (dash-dotted line) which corresponds to an atmosphere embedded in the
circumstellar disk.
As soon as the drop in density on the outer boundary starts, an outward flow of the atmospheric
gas sets in. While the density profiles follow the general outline of the sequence of stationary models
presented in Fig.~\ref{FIG_structure-rhoOBC}, they also show imprints of dynamical processes.
First, during the phase of the outer boundary density reduction (i.e., the first 32 years)
a prominent bump appears in the density profiles slowly traveling outward.
This bump marks the interface where the establishing outflow from the inner region meets
the quasi-hydrostatic outer part that is
characterized by a smooth connection to the changing outer boundary condition.
In the temperature plot in the lower panel of Fig.~\ref{FIG_M01dyn} these shocks appear as spikes,
in the corresponding plot for the 2~$M_{\rm \oplus}$ core in Fig.~\ref{FIG_M2dyn}, the spikes are too small to stand out.
As the outer, thin part of the atmosphere has a very short relaxation timescale, it
can follow almost instantaneously the density changes imposed on the boundary and
remains close to a hydrostatic stratification.

In the initial expansion phase, the launch region of the outward flow reaches into the optically thick part
where the expansion causes a temporary drop of the temperature below the level of the isothermal part,
producing prominent dips in the temperature profiles.
These temperature dips, however, do not persist and are soon filled up by radiative transport.
The Kelvin Helmholtz timescale of the relevant part of the atmospheres is about 0.3 years for the
0.1~$M_{\rm \oplus}$ core and about 10 years for the 2~$M_{\rm \oplus}$ core, i.e., significantly shorter than the dynamical outflow timescale.
At later stages in the evolution, the acceleration of the outflow mainly occurs in optical thin regions
where radiative transport is much more efficient and thus prevents the formation of these low temperature dips.

After the initial dynamical processes, a phase of free outflow follows.
As the variation
of the outer boundary density has already reached its final (i.e., nondisk) value by then, the density
profiles in this phase exhibit
a discontinuous kink at the outer boundary.
In this phase, the physical structure is determined throughout by the outflow originating in the deep atmosphere.
In the outer part of the atmosphere, the density stratification therefore closely resembles an isothermal wind.

Finally, at the end of the evolution another inward traveling bump is
produced where the dynamic outflow
again meets an outer quasi-hydrostatic region attached to the outer boundary condition.
This quasi-hydrostatic region slowly grows inward and
ultimately the dynamic outflow entirely dies away and the whole atmosphere resettles to a hydrostatic configuration.
Accordingly, the final hydrostatic configuration (dashed line) again exactly resembles the
stationary solution as illustrated in Fig.~\ref{FIG_structure-rhoOBC}.

The dynamic outflow, and in particular the bumps appearing between flowing and quasi-hydrostatic parts of the
atmosphere, should not be confused with acoustic waves. Acoustic waves naturally appear in the solution,
and the typical sound crossing time through the atmosphere is on the order of $10^6$ s.
Acoustic perturbations, however, do not contribute to mass transport and only very insignificantly
(as compared to the luminosity) to energy transport.
Since the resolution of acoustic waves is very much a numerical inconvenience
in that it requires very small time steps,
we aim to suppress acoustic perturbations in dynamic outflow simulations whenever possible (see Sect.~\ref{SECT_numericalparameters} below).

\begin{figure}
\includegraphics[height=1.\columnwidth,angle=-90]{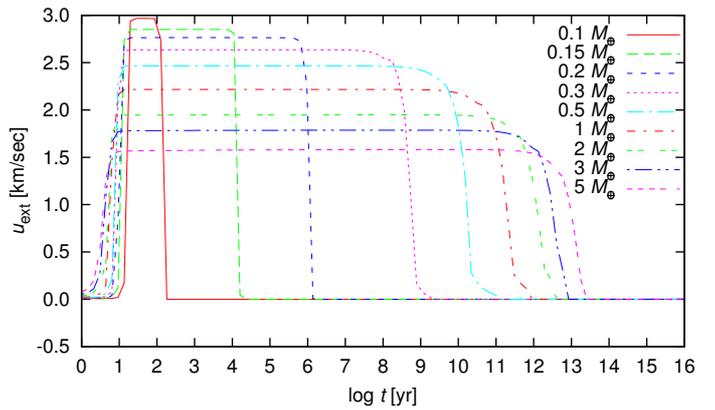}
\caption{
Outflow velocity over the outer boundary as a function of time in dynamic outflow simulations for
atmospheres around planetary cores of different mass.}
\label{FIG_u_ext}
\end{figure}

\begin{figure}
\includegraphics[height=1.\columnwidth,angle=-90]{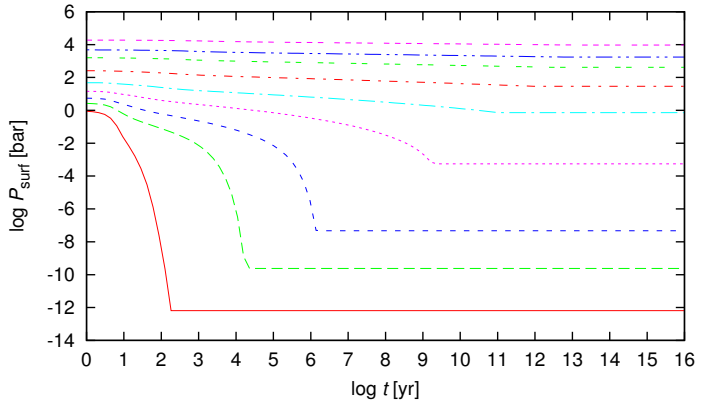}
\caption{
Atmospheric pressure at the surface of the solid core as a function of time
in the dynamic outflow simulations.
The line-type coding is the same as in Fig.~\ref{FIG_u_ext}.
}
\label{FIG_P_in}
\end{figure}

\begin{figure}
\includegraphics[height=1.\columnwidth,angle=-90]{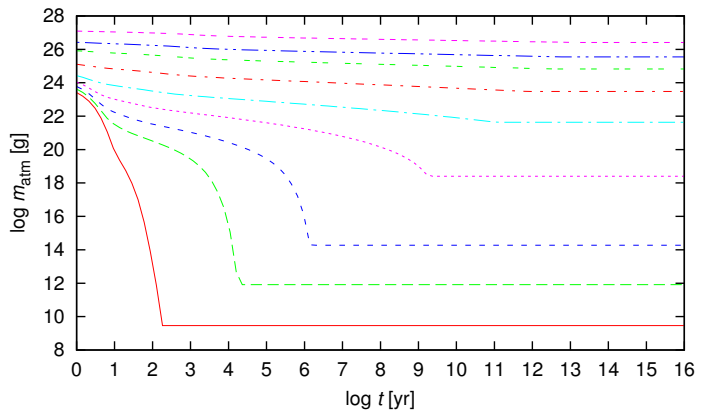}
\caption{
Total mass of the atmosphere in the dynamic outflow calculations
as a function of time for planetary cores of various masses.
The line-type coding is the same as in Fig.~\ref{FIG_u_ext}.
The different initial curvature of the lines, as compared to Fig.~\ref{FIG_P_in},
is due to the fact that the total mass not only includes
gravitationally accumulated gas but also background disk structure within the Hill sphere that can
contribute significantly in the case of thin atmospheres.
}
\label{FIG_m_atm}
\end{figure}

The timescale of the outflow and atmospheric evolution is studied in
Fig.~\ref{FIG_u_ext}, Fig.~\ref{FIG_P_in}, and Fig.~\ref{FIG_m_atm}.
The difference in timescale can be best appreciated
from Fig.~\ref{FIG_u_ext} showing the outflow velocity over the outer boundary as a function of time.
For comparison, the speed of sound in the outer atmosphere is about 0.8~km/sec.
The outflow velocity remains essentially constant during the outflow phase
and the duration of the dynamical outflow varies drastically with the mass of the solid core.
Whereas the dynamical evolution last about 100 years for an 0.1~$M_{\rm \oplus}$ core,
it already last $10^{6}$ years for an 0.2~$M_{\rm \oplus}$ core,
$10^{9}$ years for an 0.3~$M_{\rm \oplus}$ core,
and over $10^{12}$ years for cores more massive than 1~$M_{\rm \oplus}$.
Obviously, it is not sensible to interpret these long-term evolution results literally as
they are based on assumptions (e.g., accretion luminosity) that will not remain constant nearly long enough.
The results are, however, useful to study the inherent dynamical timescale of the planetary atmospheres.

The evolution of the total atmospheric mass is illustrated in Fig.~\ref{FIG_m_atm}.
A more appropriate measure for the amount of atmosphere is, as mentioned above, the pressure at the bottom of
the atmosphere, which is presented in Fig.~\ref{FIG_P_in}.

The magnitude of the evolution timescales for cores of different mass can be verified from a simple
argument based on stationary models.
First, we note that the density in the outer part of the atmosphere (not necessarily the density
boundary condition) evolves over time in a continuous way. Therefore, the evolution of the atmospheric
structure can reasonably be approximated by a sequence of stationary models with different outer densities
such as presented in Fig.~\ref{FIG_structure-rhoOBC}. The models in the sequence
possess different atmospheric masses, and in an assumed evolution this mass has to be transported over the
outer boundary. The density at the outer boundary is known from the boundary condition,
and for the flow velocity we can put for the kinetic energy in an isothermal outflow
$e_{\rm kin} = \frac{1}{2} \rho u^2 \simeq P$ and thus estimate $u \simeq c_{\rm s}$.

\begin{figure}
\includegraphics[height=1.\columnwidth,angle=-90]{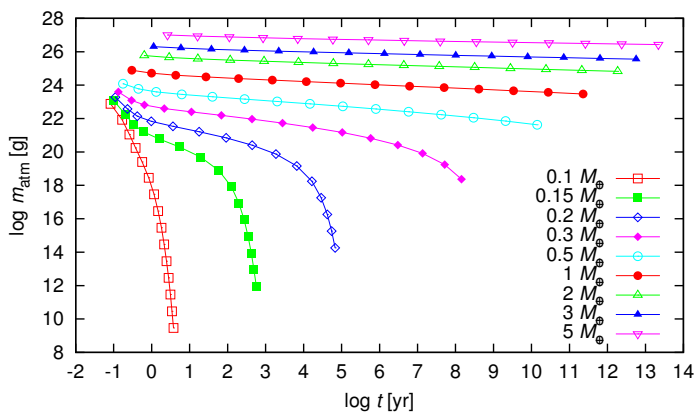}
\caption{
Total atmospheric mass of stationary models
around cores 0.1~$M_{\rm \oplus}$ and 5~$M_{\rm \oplus}$
plotted on a time axis calculated from a simple model for the transition between individual
stationary atmospheres (see text).
}
\label{FIG_m_atm_stat}
\end{figure}

Now provided with a difference in mass, a cross section (Hill sphere), a flow density, and flow velocity,
we can obtain an estimate for the evolution time to get from one stationary model to the other.
Summing up for a sequence of models yields a pseudo time-evolution of the planetary atmosphere.
These pseudo time-evolution scales for the atmospheric mass are illustrated in Fig.~\ref{FIG_m_atm_stat}
for planetary cores between 0.1~$M_{\rm \oplus}$ and 5~$M_{\rm \oplus}$.
Comparison with the equivalent numerical results in Fig.~\ref{FIG_m_atm}
reveals that the simple flow argument is able to confirm the range of
evolution timescales in the hydrodynamics simulations.
The timescales derived from the sequence of stationary models are systematically shorter
than the timescales in the dynamic models.
This, however, does not affect the basic argumentation but only reflects the simplifications of the method.
While the simple mass flow estimate may give about the right order of magnitude, it cannot be expected to reproduce the results of hydrodynamical
simulations with any accuracy.

The stark contrast in the atmospheric reaction (measured in $P_{\rm surf}$) to
the drop in outer boundary density
between cores of different mass has already been illustrated in Fig~\ref{FIG_P-rhoOBC}.
In addition, Fig.~\ref{FIG_P_in} also highlights the very different timescales on which these reactions happen:
Therefore, not only are the consequences for the atmospheric structure to
the removal of the embedding nebula gas
for a massive core much smaller than for a low mass core,
the dynamic reaction is also much slower and thus
the atmosphere of the massive core is unlikely to evolve very far in any reasonable time.

%%%%%%%%%%%%%%%%%%%%%%%%%%%%%%%%%%%%%%%%%%%%%%%%%%%%%%%%%%%%%%%%%%%%%%%%%%%%%%%%%%%%%%%%%%%%%%%%%%%%
\subsection{Physical parameters}
\label{SECT_physicalparameters}

To obtain a measure for the dependencies of the results
on both the physical and numerical parameters involved in the calculations, we investigate
here, and in Sect.~\ref{SECT_numericalparameters},
a parameter space in the most obvious and important parameters.

\begin{figure}
\includegraphics[height=1.\columnwidth,angle=-90]{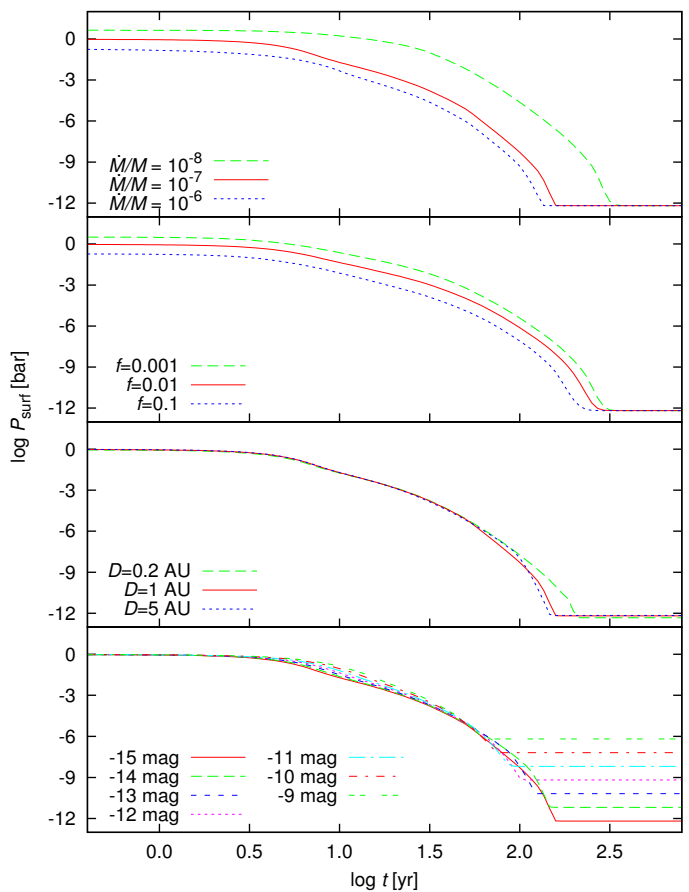}
\caption{
Effect of important physical parameters on the
surface gas pressure as a function of time
in dynamical simulations of an atmosphere around an 0.1~$M_{\rm \oplus}$ core.}
From top to bottom:
Accretion rate of planetesimals onto the planetary core (defining the planetary luminosity);
dust grain depletion factor $f$;
orbital radius of the planet;
magnitude of the drop in nebula gas density, starting in each case from $5 \times 10^{-10}$~g/cm$^3$.
\label{FIG_M01test}
\end{figure}

\begin{figure}
\includegraphics[height=1.\columnwidth,angle=-90]{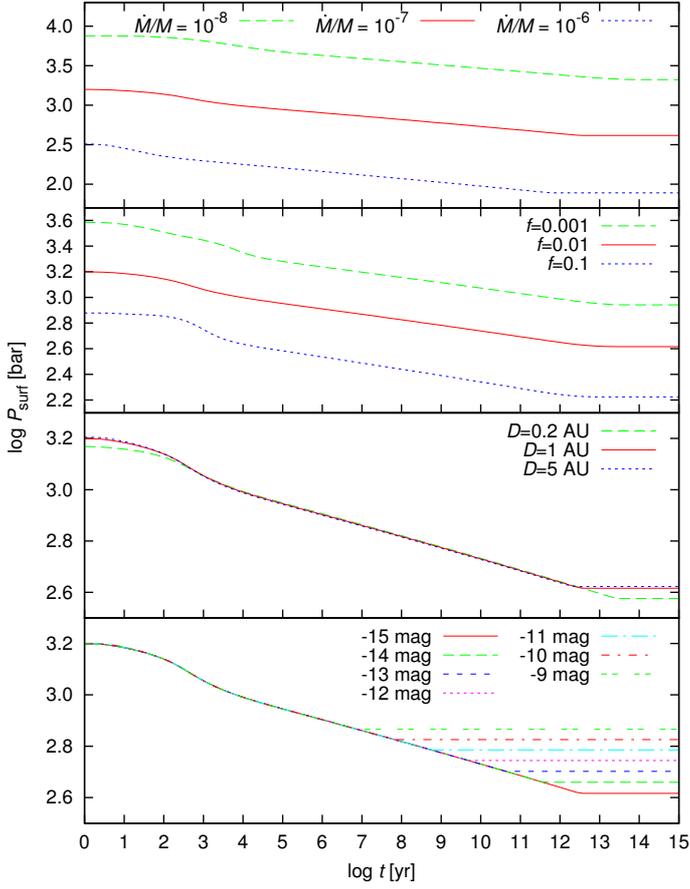}
\caption{Same as Fig.~\ref{FIG_M01test}, for a planetary core of 2~$M_{\rm \oplus}$.}
\label{FIG_M2test}
\end{figure}

Apart from the mass of the solid core,
the two most important constitutive parameters of stationary models of planetary atmospheres are the planetary luminosity, here associated with a mass accretion rate, and the dust grain depletion factor $f$.
Unfortunately, both of these parameters are very poorly constrained both from theory and observations.
The effect of these parameters is illustrated in the first and second panel of Fig.~\ref{FIG_M01test} and Fig.~\ref{FIG_M2test}, for
solid cores of 0.1~$M_{\rm \oplus}$ and 2~$M_{\rm \oplus}$, respectively.

The most obvious effect of both the planetary luminosity and the dust depletion factor is the
scaling of the amount of atmosphere in the initial models and thus of the atmospheric surface pressure.
The sensitivity to the parameters is about the same in the two examples,
even though the difference is harder to pick out in case of 0.1~$M_{\rm \oplus}$ core because of the
larger scale of the pressure axis.

The influence of the outer boundary conditions on the atmospheric models
has been tested by
varying the orbital distance and the magnitude of the drop in nebula density in the
third and fourth panel of Fig.~\ref{FIG_M01test} and Fig.~\ref{FIG_M2test}.

The orbit of the planet enters Eq.~\ref{EQ_HillRad} for Hill radius and thus defines the localization of the outer boundary of the model.
As both the orbital radius and the mass of the host star only enter in this equation,
it is sufficient to consider here the orbital distance only.
We do not take the effect that a different orbit or host star mass
will also imply a different disk gas temperature into account.
Figure \ref{FIG_M01test} and Fig.~\ref{FIG_M2test} show, however, that the placement of the outer boundary,
as specified through the orbital distance, does not appreciably affect the results.
This is reassuring insofar as the placement of the outer boundary is a somewhat arbitrary decision.

The magnitude of the density drop, i.e., the assumed remaining ambient density after the protoplanetary disk has evaporated,
only has the effect of defining the overall duration of the outflow phase.
The physical behavior and the intrinsic timescale are unchanged.
Thus, the simulations evolve along a very similar pattern but are terminated at a stage according to the final ambient density value.
In terms of outflow duration,
the effect is larger for the 2~$M_{\rm \oplus}$ core than for the 0.1~$M_{\rm \oplus}$ core
because of the different slope of the atmospheric evolution tracks.

\begin{figure}
\includegraphics[height=1.\columnwidth,angle=-90]{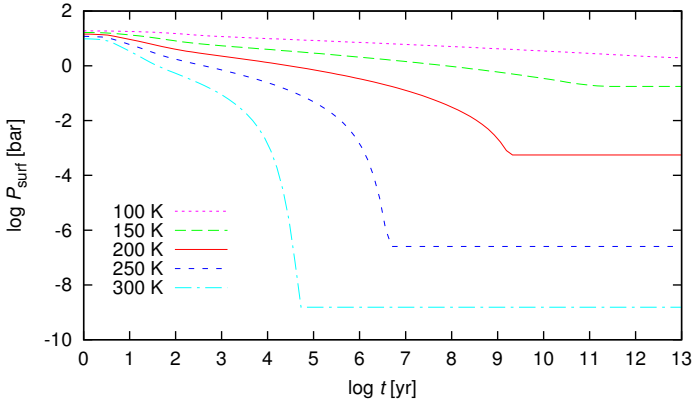}
\caption{
Atmospheric pressure at the surface of a 0.3~$M_{\rm \oplus}$ core as a function of time,
calculated for different temperatures of the surrounding nebula.
}
\label{FIG_Tdisk}
\end{figure}

Finally, in Fig.~\ref{FIG_Tdisk}, we illustrate the effect of the temperature of the disk environment
on the atmospheric structure and evolution.
In general, the lower the temperature of the embedding nebula, the more dense (and thus the more massive) is a stationary envelope.
For the two-part analytical model, this is directly evident from Eq.~\ref{EQ_outerpart} and Eq.~\ref{EQ_innerpart}.
As the amount of atmosphere is a main factor determining the time evolution in the outflow scenario,
the model series with different temperatures in Fig.~\ref{FIG_Tdisk} looks
strikingly similar to a series with varying core mass (Fig.~\ref{FIG_P_in}).
The large spread in evolution timescales in Fig.~\ref{FIG_Tdisk}, however, is much reduced
when focusing on planets in the habitable zone as this imposes strict limits on the ambient temperature.

%%%%%%%%%%%%%%%%%%%%%%%%%%%%%%%%%%%%%%%%%%%%%%%%%%%%%%%%%%%%%%%%%%%%%%%%%%%%%%%%%%%%%%%%%%%%%%%%%%%%
\subsection{Numerical parameters}
\label{SECT_numericalparameters}

Apart from physical parameters discussed above in Sect.~\ref{SECT_physicalparameters},
several purely numerical parameters affect the solution and deserve discussion.

\paragraph{Convection parameters:} The turbulent convection model involves several parameters,
most importantly the mixing length parameter $\alpha_{\rm ML}$.
The set of standard parameters,
adjusted with reference to the mixing length theory of convection,
has been compiled in Tab.~\ref{TAB_convparas}.
Overall, the effect of the convection parameters remains small.
For reasonable choices of $\alpha_{\rm ML}$, the structure is very close to the adiabatic one at all times.
Therefore, the traditional and more simple approach
to the modeling of convection zones in planetary atmospheres
of directly using the adiabatic gradient once the Schwarzschild criterion is fulfilled is well justified.
In the present study, however, this sort of approach would not be applicable as hydrodynamic simulations
also require a dynamical convection model.

\paragraph{Viscosity:} Dissipation enters the scheme in several distinctive ways: from temporal discretization,
spatial discretization, and from artificial numerical viscosity.
One particular problem in the dynamical escape simulations is the need to suppress acoustic waves, which
would travel back and forth in the thin outer atmosphere as soon as the density variation at the outer boundary begins.
These acoustic waves do not contribute to the evolution of the atmospheric structure,
and yet the small time steps necessary to resolve these waves
would make it impossible to cover the relevant evolutionary timescales in all but a few cases.
Acoustic perturbations in planetary atmospheres are subject to physical dissipation processes, such as
radiative cooling, turbulence, and non-radial dispersion.
These effects, however, are either small (radiative cooling) or not included in the model.
Therefore, depending on the particularities of the atmospheric model, sometimes significant numerical viscosities had to be deployed to raise the overall
dissipation of the code to levels high enough to suppress such acoustic waves.

So while \emph{some} dissipation remains indispensable, it is possible to disentangle the effect of the individual contributions to some degree.

First, temporal discretization errors
do not affect the solutions spatially but only dissipate fast (high frequency) physical variations.
In the application to evolution simulations, this is a very desirable effect
and thus we used a first order time integration scheme for all time-dependent calculations.
Dissipation from temporal discretization errors is the most important means to suppress acoustic perturbations and
therefore a second order scheme would only be advisable when one is interested in oscillatory phenomena.

The effect of the artificial viscosity can be studied
from the comparison of simulation runs without (for a model where this is possible) and with various amounts of viscosity.
In the production runs viscosity parameters between 0 and 1 have been used, and in this range the
viscosity only moderately affects the overall results, the largest effect being apparent for the least massive cores.
But even in the case of the 0.1~$M_{\rm \oplus}$ core, the effect is only a slight change in the exponential slopes
in the evolution of variables
with a corresponding change in
evolution time,
the solution otherwise remains very similar qualitatively.

Dissipation from spatial discretization scales with the spatial resolution
to the same order as the accuracy of the numerical scheme, which here is essentially of second order.
To estimate this effect, we changed the number of grid points from the default number of 500 in a range between
100 and 5000. Another way to reduce spatial discretization errors is to enable the adaptive grid equation, which
continuously redistributes a fixed number of grid points according to gradients in the evolving solution and
thus increases the effective spatial resolution of the physical variables.

The result of both these numerical tests is that the effect of the
spatial discretization errors on the solutions of the time-dependent calculations
is comparable to, but smaller than, the effect of artificial viscosity (in the adopted parameter range).
From this we can estimate that, for the results presented in Sect.~\ref{SECT_dynamic_results},
the inaccuracies introduced from the limited spatial resolution are
smaller than the (controlled) effect of the artificial numerical viscosity.

%%%%%%%%%%%%%%%%%%%%%%%%%%%%%%%%%%%%%%%%%%%%%%%%%%%%%%%%%%%%%%%%%%%%%%%%%%%%%%%%%%%%%%%%%%%%%%%%%%%%
\section{Discussion}
\label{SECT_Discussion}
%%%%%%%%%%%%%%%%%%%%%%%%%%%%%%%%%%%%%%%%%%%%%%%%%%%%%%%%%%%%%%%%%%%%%%%%%%%%%%%%%%%%%%%%%%%%%%%%%%%%

For the astrophysical interpretation of the numerical results,
it is important to keep in mind the assumptions and limitations inherent to the numerical scheme.
First, the assumption of a constant planetary luminosity during the whole
evolution scenario is clearly not realistic.
In the disk-phase, gas-drag is a possible way to mediate migration of planetesimals \citep{Weidenschilling_1977}
and thus to provide a resupply for the accretion of planetesimals onto the planet.
It seems likely that this changes fundamentally once the disk evaporates and
in later stages of the evolution of the planetary system, the accretion rate will soon become
insignificant for the energy balance of the planetary core.

Second, the numerical treatment of some atmospheric physics is rather simple:
chemical inhomogeneities are neglected and the simple dust model negates effects such as grain growth or atmospheric rain-out.
The radiation field, which, corresponding to the complex materials physics, will have a
very feature-rich spectral distribution,
is approximated by gray radiative transport and frequency averaged opacities.
Furthermore radiative transport is essentially diffusion-like and irradiation from the host star
onto the planetary atmosphere is not explicitly modeled.

While some of the numerical limitations might be lifted by future advances of the code,
some other restrictions, e.g., the uncertainties in the modeling of dust, seem unavoidable.

However,
it is not the scope of this paper to predict realistic evolution scenarios of specific planets,
nor to construct detailed structure models for planetary atmospheres,
but to explore the general trends and the dynamical behavior of disk-accumulated planetary protoatmospheres.
As the test-calculations against physical parameters presented in Fig.~\ref{FIG_M01test} and Fig.~\ref{FIG_M2test}
illustrate, the general physical behavior is not qualitatively affected, even if one allows for
significant variance in important parameters.
In particular, there appear to be some robust conclusions that can be drawn from the numerical results.

The stationary atmosphere models demonstrate, that
the dependency of the atmospheric structure on the surrounding environment
varies in a very nonlinear way with the mass of the solid core.
There is a distinct cutoff between low-mass cores, which lose almost all of their atmosphere once
they are removed from the disk nebula environment
and more massive cores that keep a substantial atmosphere.
The position of this cutoff in core mass, however, is subject to
physical parameters that both reflect uncertainties in modeling as well as
actual differences from one planetary system to the other.
The analytic approximation reveals that this difference
corresponds to atmospheres with deep adiabatic parts that are comparatively independent of the outer boundary conditions
and atmospheres around small cores, which are almost entirely isothermal and strongly scale with the density
at the outer boundary.

The time-dependent simulations exhibit an equally drastic difference in
the timescale of atmospheric outflow.
Low-mass planetary cores not only lose almost all of their atmosphere with the evaporation of the circumstellar disk,
they also do so very quickly compared to other evolutionary processes.
More massive cores, on the other hand, show very long evolution timescales and
it seems clear that atmospheric mass loss by dynamic outflow is not efficient for massive cores, and consequently
they will retain a large part of the atmosphere accreted in the disk-embedded phase.
The cooling of the planet in later stages of evolution and the
decreasing planetary luminosity will only enhance this trend.
Hydrodynamic atmospheric loss caused by the high XUV radiation of the protoplanet's young host star is also unlikely
to remove a significant fraction of nebula-captured hydrogen envelopes around more massive planetary cores \citep{Erkaev2013, Lammer2014}
in orbital distances related to the habitable zone of solar-like stars. Moreover, nonthermal atmospheric escape processes, such as pick up of planetary
hydrogen ions by the stellar wind, are also too weak to remove dense nebula-captured hydrogen envelopes from so-called ``super-Earths'' with
core masses above 1.5 Earth-masses \citep{Kislyakova2013, Kislyakova2014} at 1~AU.
As we have shown in Sect.~\ref{SECT_dynamic_results},
the range of evolution timescales in the hydrodynamics simulations
can also be understood from
a simple model for the transition between individual models in a sequence of stationary atmospheres.

A very general conclusion one can draw from the time-dependent simulations is that
the atmospheric relaxation timescales frequently become very long and that one cannot rely on the assumption that
a planetary atmosphere will always be close to the stationary solution.
In particular, the atmospheric relaxation timescales can also become very large with respect to the evolution and evaporation timescale of the protoplanetary disk.
Therefore, numerical modeling of disk-accreted planetary atmospheres should not only
rely on stationary models but be based on time-dependent methods.

\begin{acknowledgements}
The authors acknowledge the support by the FWF NFN project S116 ``Pathways to Habitability:
From Disks to Active Stars, Planets and Life'', and the related FWF NFN subprojects, S 116 02-N16 ``Hydrodynamics in Young Star-Disk Systems''
and S116 07-N16 ``Particle/Radiative Interactions with Upper Atmospheres of Planetary Bodies Under Extreme Stellar Conditions''.
\end{acknowledgements}

\bibliographystyle{aa}    % style aa.bst
\bibliography{planetenrefs}

\end{document}